\documentclass[copyright]{eptcs}
\usepackage[dvips]{graphics}
\providecommand{\event}{}
\providecommand{\volume}{2}
\providecommand{\anno}{2009}
\providecommand{\firstpage}{15}
\usepackage{breakurl}

\title{Service-oriented Context-aware Framework}
\author{
L{\'a}szl{\'o} Kov{\'a}cs
\qquad
  P{\'e}ter M{\'a}t{\'e}telki
\qquad
  Bal{\'a}zs Pataki
  \institute{MTA SZTAKI} 
  \institute{Computer and Automation Research Institute of the Hungarian Academy of Sciences\\
  Department of Distributed Systems\\
  1111 Budapest, Kende u.~13-17, Hungary}
  \email{\{laszlo.kovacs,peter.matetelki,balazs.pataki\}@sztaki.hu}
}
\def\titlerunning{Service-oriented Context-aware Framework}
\def\authorrunning{L.~Kov{\'a}cs \& P.~M{\'a}t{\'e}telki \& B.~Pataki}

\begin{document}
\maketitle

\begin{abstract}
Location- and context-aware services are emerging technologies in mobile and desktop environments, however, most of them are difficult to use and do not seem to be beneficial enough. Our research focuses on designing and creating a service-oriented framework that helps location- and context-aware, client-service type application development and use. Location information is combined with other contexts such as the users' history, preferences and disabilities. The framework also handles the spatial model of the environment (e.g.~map of a room or a building) as a context. The framework is built on a semantic backend where the ontologies are represented using the OWL description language. The use of ontologies enables the framework to run inference tasks and to easily adapt to new context types. The framework contains a compatibility layer for positioning devices, which hides the technical differences of positioning technologies and enables the combination of location data of various sources.
\end{abstract}


\section{Introduction}

In the past years mobile and desktop computer application developers realized the need to apply location- and context data so as to help users with such awareness information. However, location data has mainly spread in the field of mobile applications while context-sensitivity is used in desktop applications. The real advantage can result from the combination of these: to create location- and context-aware services and applications. Comparing a simple location-aware audio-guide to a context-sensitive and location-aware audio guide in a museum (that is capable of ranking the showpieces and change the granularity of the presented information according to the users preference and behavior) or a simple file- and application manager on a laptop computer to a location-aware one (that displays different folders and opens specific applications in case the user resides at home or at work) we can feel the benefit of the additional location- and context information.

Apart from navigation, location-based services for mobile devices have not managed to result in such a breakthrough, like e.g.~short messaging service. Two major shortcomings for the current location-based services have been indentified. On one hand, using only location as a context does not seem to be enough to satisfy the users' augmenting needs. On the other hand, using a certain location-aware service often requires the application of a specific device that belongs to a specific positioning technology. In our research we designed and created a framework for context-sensitive service and application development that aims to solve the abovementioned problems, meaning that the framework makes it easy to use and develop smart location- and context-aware services. It even goes beyond this goal by offering a semantic basis for the information and by integrating the geometry of the environment.

In this paper we first present our motivation and goals. The following chapter gives an overview of the designed system and we also get into the details about certain important concepts and building blocks of the framework. The last chapter summarizes our work.

From the point of mobile applications, considering characteristics of the users other than their location can enrich services. This additional information can include the users' schedule, tasks, presence, preferences, age, mental and physical state (e.g.~disabilities), etc. A system can be even more circumspect by not only using the users' characteristics but also their circumstances, e.g.~the users' environment at the current location such as streets, buildings, rooms and objects. Accordingly, the model of the environment needs to be taken into account. The geographical model of the physical space should also describe the interrelations of the objects in the physical space~\cite{5}. The ability of inferencing on the environmental model is necessary in case of requests that rely on the user's real surroundings like in case of a navigation application or museum guide. By being aware of the characteristics and circumstances of the users in addition to their location, the answer to their questions can become more appropriate and can specifically cover their needs. 

The major problem when using a certain location-based service is that users are required to apply a specific positioning technology, and own a specific type of device. This prevents the smooth spread of location-based services and even makes it impossible to use them with future, not-yet existing positioning techniques. A positioning generalization layer is needed for the location-aware services to hide the differences of the numerous positioning technologies. A subsystem of this kind, a location-manager layer that joins the diverse positioning subsystems enables users and developers to create and use the positioning services irrespective of the type of the positioning devices and technologies. In addition, for users having multiple positioning devices, location data generated by these sources can be combined by the location-master layer so the precision and trustworthiness of the localization and the coverage of the positioning service can be improved. Users can also switch any time from one positioning device to another without interrupting the positioning service.

As users need intelligent services, ones that are able to answer questions that are closer to human way of thinking, much can be done to achieve this goal: more information can be used about the users themselves and their environment and more intelligent answering engines can be applied. The more context parameters are considered for information retrieval, the more beneficial and satisfactory the services will be for the users~\cite{4,5}. A system of this kind accumulates much information and interrelations, and contains a remarkable knowledge base. To insure information reuse, machine-processability and data consistency, the data must have semantic interpretation. 

This kind of a context-aware system uses many aspects (characteristics and circumstances) of the user and much information about the environment. As there is no limit for the number of types of contexts that can be helpful and applicable for the system, it is important to have a flexible data model and a clearly understandable dataset at all times. Creating rules and running automatic consistency checking can facilitate this task. Well-defined relations between the data classes and instances can be used to reveal hidden connections in the database. Using ontologies as the base of the context-aware system can satisfy all the above requirements and wishes. It also makes it possible to perform inference tasks on the data.

Our framework was designed and developed in line with the above principles so as to solve the mentioned problems. To build such an infrastructure, two levels are involved~\cite{7}: we created the conceptual context model that describes the concepts and their interrelations, and a SOA style architecture that defines the modules, their hierarchy and connections. By providing the appropriate semantic context-based services it can help to solve more general and more realistic challenges. Such a system with a semantic background, with the ability to take into account both the location and diverse context information about the user and the ability to use the environmental model for geometric calculations, is capable of responding to high abstraction level questions that are closer to human thinking, people's everyday needs. 

Although, the framework's primary target is to support application and services in the world of mobile technologies, our goal was to create a system that can be at hand for both mobile and desktop-style environments.

\section{Related Work}

One possible method to create a context-aware system~\cite{2,3} is to combine an existing information system (to access information about the users) with an awareness system (that helps to distribute context information among users' devices). The HIPPIE~\cite{4} information system can support users with location-based services during the preparation, execution and evaluation of a museum or fair visit. It takes into account the users' positions and characteristics. It is a well-designed system for single-user context support, however, is not capable of handling user interactions. Following the above idea, the authors of Awareness in Context-Aware Information Systems~\cite{5} combine HIPPIE with an event based awareness environment, ENI~\cite{6}, which is able to distribute context information to users. They outline a complex system to support awareness for nomadic users about other users residing in a similar electronic or spatial context. The combined context-aware information system, featuring user tracking in the physical world and electronic space, user modeling and an event-based awareness environment, is able to support a joint experience for local visitors and remote electronic visitors at an exhibition. Although, this project takes into account the physical location, environment and other aspects of the user, it lacks the semantic description of the information, the ability to execute inference tasks and also a sophisticated positioning manager layer that we feel necessary to create an advanced and flexible location- and context-aware framework.  

Moreover, semantic-rich approaches also exist to design a context-aware system, for example the OCCA~\cite{7} that proposes a new semantic rich context-modeling concept for collaborative environments. OCCA uses OWL~\cite{16} to describe its base ontology that the architecture builds on. In the TEAM~\cite{24} project, a similar recent EU project, an ontology-based contextual framework~\cite{22} is described to capture, access and share software developersŐ experiences, the information that usually gets lost among distributed teams. Compared to our project the main difference is the moment when context fusion is achieved: this system combines contexts asynchronously while sensing the usersŐ local environment to find out his activity. On the contrary, our system combines the contexts on-demand at the time a certain intelligent service is invoked by a user. Thanks to the semantic-rich model, it provides the possibility to infer on the knowledge base of contexts, but as this work focuses on contexts of the collaborative environments, especially on internal contexts to describe humans and tasks, integrating and using the geometry-based model of the environment are not possible within this approach. 

The literature describes other similar ontologies, such as CONON~\cite{14}, CoBrA-Ont~\cite{17} and context ontology~\cite{19}, all described in OWL. These ontologies represent the ubiquitous computing domain and consider concepts like location, time, people and devices. The advantage of these ontologies over OCCA is that they consider the physical context of the users. The main focus of these projects~\cite{14,19} differs from ours as they aim to create ontologies for context-representation, but do not intend to build a framework for creating location- and context-aware services or applications based on the semantic backend. CoBrA~\cite{18} is an advanced complex architecture to support context-aware systems, which takes into account many aspects of the user and environment. In addition to its functionality our framework provides the geometric environmental calculations and a positioning generalization layer.

Du and Wang~\cite{8} have developed a system for facilitating context-aware application development for mobile devices. Although, the basic idea may seem parallel to ours, the focus points are different. Their project focuses on providing a development environment with code generation tools to create applications, while our research focuses on creating a framework providing location- and context-aware services for developers to create applications.

Devaraju el al.~\cite{9} have worked on a context gathering framework to support context-aware mobile solutions that deals with the heterogeneous data provided by sensors. This research describes in more detail what we call ``Context Middleware'' and ``Contexts'' tiers in our architecture but omits the higher-level logical layer, our ``Core'' of the framework.

In 2006, when this research started the solution to combine the various positioning technologies corresponding to a single user was a novel one. Nowadays, we can find some applications, e.g.~Google Maps~\cite{11} using the Google My Location API~\cite{12} that benefit from a similar idea. Although, Google Maps can use various positioning technologies to find the users' current location, it does not combine the information from the different sources.

In environments where high precision localization is needed but not possible, an image or video of the local environment can help positioning. Inspired by the ideas~\cite{23} presented in MOBVIS~\cite{25}, a recent EU project, low-precision location information can be combined with object recognition of the local visual information and result in high-precision localization.

The above evaluation of related projects show that, although, the low level details of our concept of a location- and context-aware framework are known to researchers, there has not been a project that combines all the essential necessities - the geometric model of the environment, the location information together with other contexts of the user and the semantic representation of all the above information -- all of which we believe to be necessary to create smart semantic-rich context-, location- and environment-aware services. In addition to the previous combination of system components, we also provide a positioning generalization and arbitration module to support users carrying and using several types of positioning devices to increase the coverage and accuracy of the location-aware services.

\section{The Framework}

\begin{figure}
\center
\resizebox{16cm}{!}{\includegraphics{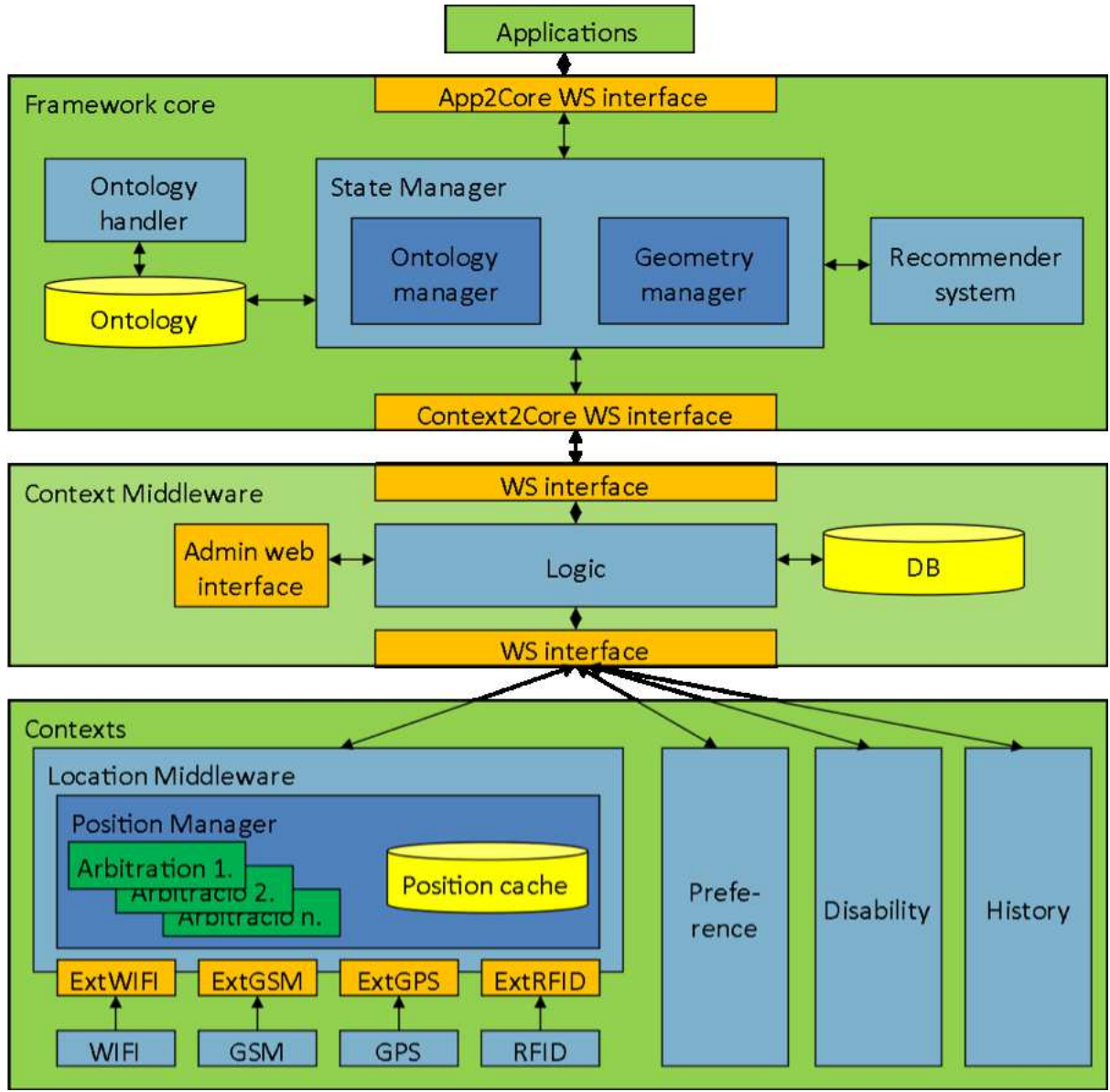}}
\caption{The components of the framework.}
\label{fig:1}
\end{figure}

Figure \ref{fig:1} shows the architecture of the framework. 

Following the paradigm of service-orientation, the functions of the system are separated into distinct units that are accessible as services over the network.

Contexts can be found in the lowest tier in the framework that contains the hardware and software components for context perception. The collected context values are forwarded to the Framework core. Intelligent contexts (e.g.~location) can perform the framework's commands or reply its requests, e.g.~to send context information continuously, on demand or in bursts. In the context-aware system we encounter several types of contexts, either simple or complex ones. The contexts in this tier may be inhomogeneous, requiring different interpretation and handling. The following list shows examples of possible contexts:
\begin{itemize}
\item Presence: is the user available at the given moment? Is a certain context of the user available at the moment?
\item Time: it can be local time (hours and minutes) or the current season, decade, etc.
\item Means of transportation: on foot, by car, using public transport
\item Social issues or status: is the user a child, a pensioner, a worker?
\item Role: can be the users social role (grandfather) or the role held in an office (manager)
\item Operating system: the contexts of the device used are also contexts of the user
\item Expertise or qualifications: people having different qualifications require different information, so this defines the style or granularity of the information
\item Coordinates, velocity: for navigation applications for drivers and excursionists
\item Traffic: dynamic traffic information for vehicle-oriented navigation
\item Environmental model: countries, cities, streets, houses and even furniture
\item Activity: what files are open/edited by the user, what is his current schedule is, where is he currently walking/driving, etc.
\item Preferences, taste: ``above 300HP'', ``Greek mythology'' or ``cheapest route'' are all types of preferences. This category can involve a very wide variety of information. 
\item History: a list of the past actions of the users, similar to detailed activity log in~\cite{5}. In case the users rate the system's answers according to their satisfaction a recommender system can enhance the framework to be an adaptive, learning one.
\end{itemize}

The location context is a complex subsystem managed by the Location Middleware. It contains all the different, non-compatible positioning technologies and masks the differences so the framework can integrate any kind of positioning system via the Location Middleware. This module enables users to profit from the simultaneous or alternating usage of their different positioning devices. In an area covered with RFID readers and GSM signal, high precision global positioning is impossible with the existing tools. However, by using a simple mobile phone and an RFID tag our system is able to determine the global position with high accuracy although neither the GSM nor the RFID technology is capable of doing this by itself. Using this module it is possible to achieve an increased precision in localization by automatically combining the sources of different positioning devices. Another benefit of this module is the possibility to detect the users' location in changing conditions, for users switching from one positioning device to another (e.g.~caused by signal loss). This is useful in scenarios where both indoor and outdoor positioning is needed. Outside the user can be positioned by his GPS while getting indoors (and losing the GPS signal) the positioning continues as the GSM or with other indoor positioning systems that become accessible. 

Our system was designed in a way that is able to represent location information acquired from various positioning sources. This ensures the compatibility with existing and future positioning solutions. Within our project two scientific groups are researching alternative positioning technologies and building prototype systems. One group for RFID and the other one for WLAN based positioning. We cooperated with these teams to inspect their solutions and to integrate the positioning solutions into the framework. For the position data input we provide a Web Service that accepts location data input using a fixed parameter set. Apart from raw coordinate data, location data also includes the coordinate system, precision and probability information. So as to accommodate to the model of the framework, each team developed an extension to their system (Ext\_RFID and Ext\_WLAN according to the previous sections) that sends the position information to the Web Service. 

Apart from the location, other contexts can serve as input of the system, e.g.~time, user preferences, disabilities and history. Every context has its ontology model in the framework. The storage and querying of the heterogeneous contexts are executed through the Context Middleware using Web Service interfaces.

History is a special type of context. It does not provide fresh information but stores old information as historic data for latter use as time-stamped context-states. It's a repository-like context for storing data to provide querying possibilities later on. The information found in the History can be used by a recommender system.

The goal of the Context Middleware tier between the Framework core and the contexts is to mask the differences between the different context types and allow unified context handling and querying for the Framework core. This module transforms the heterogeneous context data into a homogeneous, semantic representation enabling the Framework core to operate independently from the structure of the different contexts and ensuring that the system can easily integrate future contexts. It also forwards the framework's requests to the intelligent, interactive contexts. The Context Middleware contains a database that matches the context-devices with their owner (a user in the system) and with the supported contexts. This database also stores other device- and context related information. As the Context Middleware hides the low-level devices from the upper tiers, it enables the Framework core to work with high-level user-assigned context data so it does not have to deal with low-level devices.

The Framework core between the contexts and the applications realizes the main functionality. This is the core of the framework. It accepts requests from the Applications, evaluates them according to the information provided by the contexts. The semantics of the context data is contained by the ontology. The main module of the core is the State Manager that can be queried for the actual state or history of the contexts. The response is provided to the applications by the Application2Core interface. 

The Context2Core Web Service interface accepts asynchronously incoming context data from the Context Middleware. It can also work as a command-generator: in case the State Manager is in need of fresh data input it requests context data from the Context Middleware in burst or single input mode through this interface. The gathered context-information is forwarded to the State Manager for further processing and storage.

The ontologies provide the semantic and logical model of the contexts. Every context has a mapping and description in the ontology. Employing semantic-aware components like ontologies in a complex system is beneficial in a number of ways. Ontology is used to describe vocabularies, machine-interpretable definitions of concepts and relations~\cite{13} between the individuals of the domain: it makes sense of the things. In machine-to-machine interactions, just like in human interactions, it is important that each party shares the same understanding of the structure and meaning of the exchanged information. This is one of the main reasons to develop ontology-based data models~\cite{13,14} in addition to enabling the reuse of the knowledge and enabling the use of reasoning engines to infer on the data. Ontologies have been used in numerous other research projects~\cite{14,18,19} to represent contextual information.

\begin{figure}
\center
\resizebox{16cm}{!}{\includegraphics{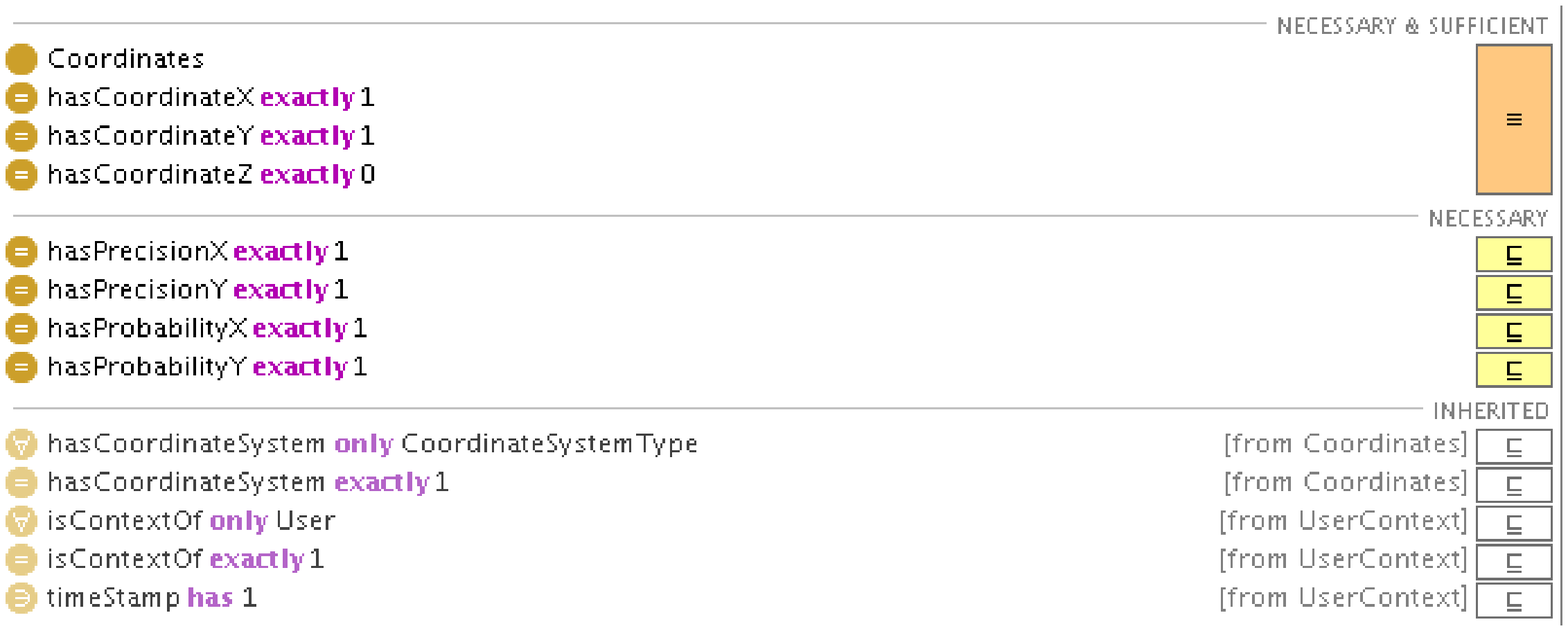}}
\caption{The definition of Coordinates2D class.}
\label{fig:2}
\end{figure}

In our ontology the so-called UserContext class is the main class for the miscellaneous context information. All context data must correspond to one exact user and must have a timestamp. As described before, many kinds of context information can be possibly added to the system. As an example disabilities, preferences and location were defined. Our Coordinates class must have exactly one coordinate system from the CoordinateSystemType instances. As depicted on figure \ref{fig:2}, a two dimensional coordinate must have, in addition to the previous constraint, probability and precision values in the X and Y axes, coordinate values in X and Y directions but must not have a Z coordinate.

To describe the geometric shapes we use the W3C Geospatial Vocabulary~\cite{21}, created by the W3C Geospatial Incubator Group in 2007. This ontology describes the location and geometric properties of resources as ``geographic features''. Our complex geometry shapes are composed of multiple geographic features. 
The ontology handler helps to manage existing ontologies and to define and integrate new ontologies to the framework. It also functions as a browsing and querying interface for the stored instances inside the ontology.

The State Manager represents the state of the world from the point of view of the framework: it contains fresh and historical context-information about users and the spatial model of their environment. This world-model is continuously updated by the contexts. The great majority of the requests from the applications and services arrive to the State Manager that processes and answers them. So as to answer complex requests it divides them into more simple ones and forwards them to the Ontology- and Geometry managers. Gathering and combining the responses produce the final result. To enrich the high abstraction level functionality and the novelty of the framework, the mutual matching of the geometry objects to the instances stored in the ontology is very important. The Geometry manager module of the State manager uses the geometry object definitions stored in the ontology and the visually portrayed geometry objects can be associated to the corresponding ontology instances. 

The world model, such as the spatial description and all context information about the users, is semantic in our framework. The Ontology manager stores and queries the context data in the ontology. It contains an inference engine that allows executing complex SPARQL queries, revealing consistency problems and running other inference tasks on the ontology.

The ability of making calculations and inference in finite time on the context-data was a high priority when selecting the ontology description language, OWL DL~\cite{10}. As the inferencing solutions for description logic languages are not appropriate to calculate spatial relationships (defined by coordinates), a separate module is created in the framework dealing with spatial calculus: the Geometry manager. As part of the State manager this module executes tasks related to spatial calculations. The spatial model of the environment can be represented in 2D or 3D. The system also provides the possibility of importing the models and worlds described in 3D modeling languages (e.g.~VRML) to the internal representation. The Geometry manager is capable of working with local and global coordinates.

Based on the users actions in the past the integrated Recommender Systems can make recommendations and create predictions. These can be used to auto-personalize the system for the users.

Application2Core interface is the Web Service interface used to communicate with applications.

The applications are not part of the framework; they are the interfaces towards the users. They can use and benefit from the simple and complex services offered by the framework through its Web Services. Applications may be as simple as a user tracking software or may be complex, intelligent applications using a complex service of the framework. A fat-client style complex application can use the services of the framework, can recompose or combine them and create new services and functionality. We can imagine applications running on mobile devices, desktop or notebook PC-s or even web applications. Because the framework is designed as a generic context-support system, is can meet the needs of many sorts of applications.

\section{Sample Services and Utilization Examples}

The current version of the framework uses the relational data store of the Context Middleware for storing device related information. For storing and searching user-contexts it uses the ontology based persistence modules in the Framework core. It also has a spatial reasoning module. Combining all these modules makes the framework able to respond to complex questions where location, context (e.g.~history, preferences, disabilities) and the geometry of the environment all play an important role. We developed some sample services to demonstrate the available functionality of the framework to reveal some possible usages. 

A service where location and geometry play the most important role is the ``Can user A see user B?'' service. Positions of both users are queried from the framework and the geometry (currently in 2D) is examined for objects between the users. For a query to this service a user is given a textual answer and a graphical answer with the map of the scene and the users. In case there is an object between the users, it is highlighted with red. The textual answer also gives information about which geometry object (which geography feature and which spatial object) is between the users. 

Visibility itself is an interesting question; a detailed and accurate model and a geometry engine are needed to execute this kind of calculation. This service also proves that by combining a few contexts (the location and the environment) interesting new services can be created.  In our current sample service we could also benefit from the users' preferences or disabilities to personalize the visualization; e.g.~for people with low vision the system could generate a high contrast image.

We also created a simple service that can be used to acquire a user's location. This service has multiple purposes; on one hand, it provides an easy solution for monitoring the user's real-time location using the framework. On the other hand, service developers can use this service to create third-party location-based services based on the framework, yet those are not implemented in the framework. To demonstrate the use of this service, a dynamic webpage that shows a user's alternating location in real time was developed. The page also shows the old positions and differentiates their visualizations from the most recent location. For the real-time visualization we use the DWR (Direct Web Remoting)~\cite{20} Reverse Ajax technique.

Also, some base-level services were created for determining the users' distance from a given point, a geometric shape or complex shape in addition to services calculating the containment-relations of points, shapes and complex shapes.

As an example for complex services, we developed a distance-sensitive service that detects whenever a user's proximity -- in relation to a point, shape or complex shape -- falls below a threshold. The service can perform different operations, such as opening a file or sending a (email, IM, SMS) message. This service can facilitate a conference by automatically opening the lecturer's presentation who approaches the lectern. It can also act as an alarm-system by automatically sending an on- or offline warning (IM, email, SMS) in case someone gets too close to a certain location or detects theft of (an RFID tag equipped) object.

The Location Middleware can be used for tracking a user's position that has multiple positioning devices. It works when the technological circumstances of the positioning change, e.g.~switch from GPS to RFID. The use-case of tracking such a user with the aid of our system is as follows. A user's position is tracked by his positioning devices such as GPS, cell phone (GSM or 3G) and an RFID device. The location-data is forwarded to the Location Middleware that makes the positioning technology transparent for the upper level systems. The navigation system requests the actual position of the user from the framework, so the technological details and actual devices remain transparent to the navigation system. While the user drives to work the framework receives and combines the GPS and GSM based position information. As the GPS coordinates are of higher accuracy, its role will be more important during the calculation. While the user drives in a tunnel (where he loses the GPS signal) the position can be still determined using his mobile phone. As soon as he gets into the coverage of a WLAN (in addition to the GSM), the system will combine the different position information and create a result that is even more accurate and trustworthy. If the user enters a building that is not GPS covered but is equipped with WLAN and RFID positioning systems his position may be determined even in 3D. Neither the user nor the developer has to worry about the positioning technologies and devices as the Location Middleware masks them from the application.

The framework supports context-aware services that depend on multiple circumstances of the user. Therefore, not only the location but also other important contexts of the user is used by the system such as the disabilities and preferences to be able to provide personalized answers for the users' queries. In case a user has low vision the user interface of a service on his mobile device can display high contrast colors. In case he has mobility impairment a service helping the orientation in a museum will offer the elevator in place of the stairs. In case the user prefers the ancient aged show pieces to the new aged ones the system will give more detailed information about the preferred ones during guidance. These examples show that the benefit of using multiple user contexts helps to fine tune the service.

As an example for showing how to use the geometry of the environment of a user we can think of an intelligent museum guide. In case a visitor loses his child in the museum he can ask the intelligent guide how to find him. Beyond telling the room number the framework using the spatial model of the museum can show the route to get to the desired location. The route can be simply drawn on a graphical map and sent to the users device but as the system knows the meaning (semantics) of the geometry objects it can guide the user by telling ``\ldots enter the door on your right and turn right \ldots'' and so on.

Objects can have disqualifications for certain disabilities: an elevator meets the needs of a handicapped person but all the stairs are disqualified for this kind of disability according to the semantics of the geometry model. As soon as the disability context is also combined with the geometry and location contexts for the route planning, the framework can avoid making wrong suggestions such as using the stairs while being in a wheelchair. In case a museum visitor is blear-eyed and can see clearly only 1 meter away, the system will send the information about a showpiece to his handheld device instead of displaying it on the fixed monitor, which he would be unable to read.

In a museum environment, the use of the recommender system enables the framework to suggest unvisited showpieces for a user. Suggestions can be made based on the user's profile or based on similar tracks of other users.

\section{Summary}

Considering the present deficiencies and future expectations of location-based services a system for context- and location-aware services was designed. The framework generalizes the location-based services to context-based services; its aim is to assist context-aware services and applications in mobile or desktop environments. The framework supports location-based services, and also implements context assistance in general where users' characteristics (e.g.~disabilities and preferences) and circumstances (e.g.~the environment and history of past actions) are also part of the context. Context modeling is achieved by using the OWL description language, we use ontologies to define and store all context data. This semantic description of the data allows the system to perform reasoning and automatic consistency checking in the knowledge base, makes the data machine-processable and also provides us a flexible and expandable data model. 

Our framework also offers a subsystem masking the different positioning technologies from the users and developers. Users can use location-based services independent of what kind of positioning device they own, developers can create location-based services independent from what kind of positioning device produces the location-data input. This subsystem is able to combine the location information provided by different positioning sources to spread the positioning coverage and to help achieve a higher positioning availability and precision. 

The development of the framework is currently under process. The core parts have been created: the ontology itself, the Ontology manager and Ontology manager of the State Manager, the Recommender system and the Web Service interfaces to communicate with the applications and the contexts. The Context Middleware relational database and web interface have also been developed. To accumulate real context-data we have integrated our system with both the RFID and WLAN based positioning solutions. For testing, the sample services described in the previous sections have been used. There are two major directions to continue the development and evaluation of the system. So as to improve the functionality we can integrate new contexts by creating new contexts or finding new context providers or we can build new applications to combine the existing contexts. To specify what contexts are relevant for specific application types a sociological survey would help further research. 

The framework can provide a good opportunity in several fields in the IT sector: data providers who possess context data, a large amount of information about the users, can combine the various raw data to increase their knowledge base and to obtain additional information. Developers may also realize that combining the different contexts of the users can result in interesting applications. These and other new possibilities may become accessible via the use of the framework. With such context-aware services accessibility issues in general can be assisted. Examples may vary since not only physical disability can be taken into account. Also, the system detecting poor mobile phone signal reception can switch to VoIP communication, and the system being aware that the user is carrying heavy luggage can advice to take the escalator in place of the stairs.

\section*{Acknowledgments}
This work has been supported by the BREIN project~\cite{15} and the MIK project~\cite{1}. BREIN is partly funded by the European Commission under contract FP6-034556. MIK is funded by the National Office for Research and Technology in Hungary.

\bibliographystyle{eptcs}

\end{document}